# Modeling the cooldown of cryocooler conduction-cooled devices

Ram C. Dhuley

Fermi National Accelerator Laboratory, Batavia, IL 60510, USA

Email: rdhuley@fnal.gov

*Abstract.* Cryocooler conduction cooled devices can experience significant cooldown time due to lower available cooling capacity compares to convection cooled devices. Therefore, the cooldown time is an important design parameter for conduction cooled devices. This article introduces a framework developed in Python for calculating the cooldown profiles and cooldown time of cryocooler conduction-cooled devices such as superconducting magnets and accelerator cavities. The cooldown time estimation problem is essentially a system of ordinary first-order differential equations comprising the material properties (temperature dependent thermal conductivity and specific heat capacity) of the components intertwined with the prevailing heat transfer channels (conduction, radiation, and heat flow across pressed contacts) and the cryocooler capacity. The formulation of this ODE system is first presented. This ODE system is then solved using the in-built Python library *odeint*. A case study is presented comprising a small cryocooler conduction-cooled copper stabilized niobium-titanium magnet. The case study is supplemented with the Python script enabling the reader to simply tweak the device design parameters and optimize the design from the point of view of slow/fast cooldown.

## 1. Introduction

Cryocooler conduction-cooled devices operating near 4 K such as small superconducting magnets [1] and accelerator cavities [2-7] are fast gaining popularity as an alternative to the devices convectively cooled by liquid or supercritical helium. Such devices are inherently safer to build and operate as they lack liquid/supercritical helium around their insulation vacuum spaces, which drastically mitigates the dangers associated with a sudden loss of this vacuum [8-11]. Due to small cooling capacity of the cryocooler, however, conduction-cooled devices can have very long periods of cooldown from room temperature to the base temperature near 4 K. The design of thermal conduction links between the cryocooler and the device, therefore, needs to consider the device cooldown time in addition to the system performance at the base temperature.

FERMILAB-TM-2782-TD

This note introduces a script written in Python that enables visualization of the cooldown profile (temperature *vs.* time at various physical locations of the system) and estimation of cooldown time for cryocooler conduction-cooled devices. The inputs are cryocooler capacity (generally provided by a cryocooler manufacturer) and thermal properties of material used to construct the device (usually obtained from the NIST cryogenic material properties database [12], literature on thermal contacts [13,14], and other sources [15]). A system of ODEs comprising time dependent heat transfer equations (with conduction, radiation, contact heat transfer terms) is first formulated and then solved using the Python *odeint* library [16].

2. Methodology for estimating cooldown profile and time

The flowchart in figure 1 graphically represents the methodology. The process starts with documenting temperature dependent material properties (thermal conductivity, specific heat capacity, emissivity, and contact resistance), followed by formulation of the time dependent heat transfer equations (a system of first-order ODEs), then writing down the conduction and radiation heat transfer terms as functions of temperature, and finally solving the system of ODE.

Further details of this methodology is provided in following case study by applying it to a small superconducting magnet conduction-cooled to a two-stage cryocooler.

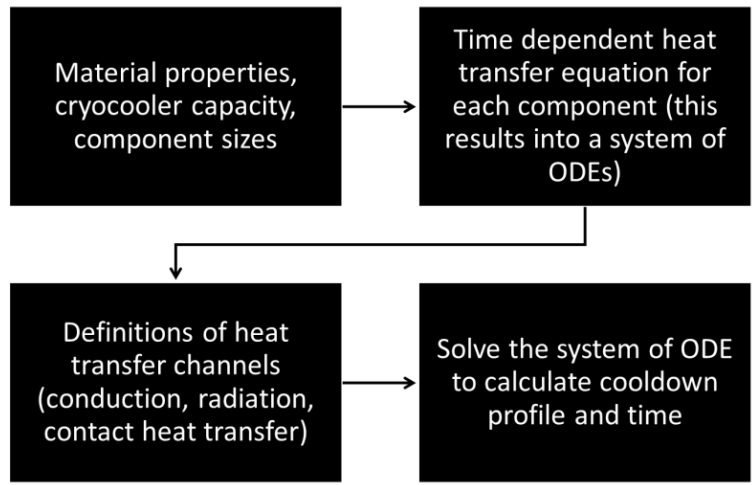

Figure 1. A flowchart for calculating cooldown profile and time of cryocooler conduction-cooled devices.

3. Case study

The methodology presented in the preceding section is illustrated using a small conduction-cooled superconducting magnet. The magnet is a solenoid with a warm bore, wound using copper stabilized niobium-titanium superconductor. The solenoid is mechanically coupled using copper thermal links to the 4 K stages of two-stage pulse tube cryocoolers. The solenoid is enclosed in an aluminum thermal shield that is conduction-cooled to the 45 K stages of the same two-stage cryocoolers. The thermal shield is wrapped with multilayer insulation to reduce ambient radiation heat leak. The entire cold is enclosed in a vacuum chamber to cut down convection heat transfer between the two. More details of the magnet system under consideration can be found in [17, 18].

The magnet system can be represented as a nodal network comprising the system components (magnet, thermal shield, cryocoolers, vacuum vessel, etc.) coupled with the heat transfer channels of conduction, radiation, and contact resistance. This is detailed in figure 2.

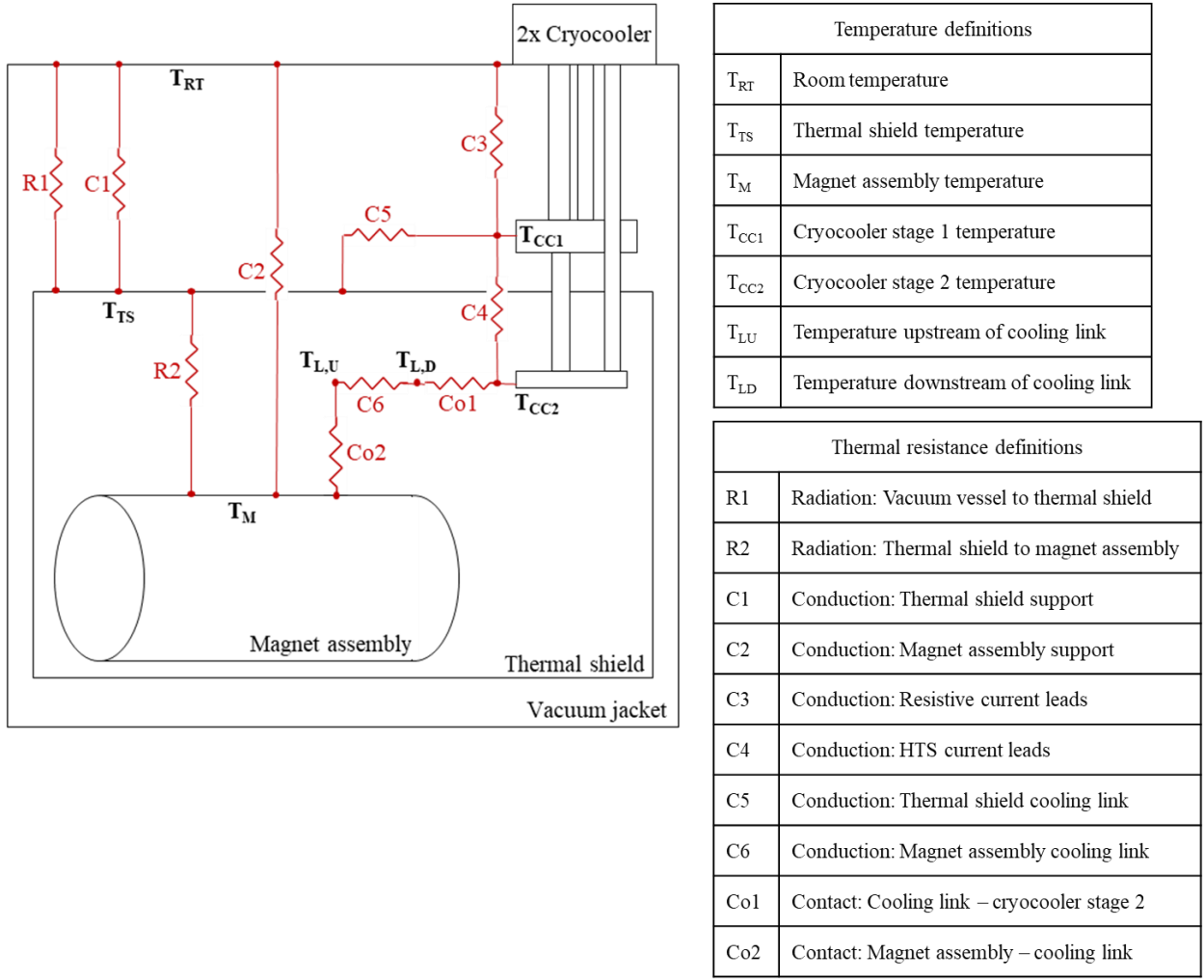

Figure 2. Nodal network and heat transfer channels for the magnet system under consideration.

The conduction, radiation, and contact heat transfer channels between any two nodes at temperature $T_1$ and $T_2$ are given by:

$$q_{\text{conduction}}(T_1, T_2) = \frac{A_{\text{CS}}}{L} \int_{T_2}^{T_1} k(T)\, dT, \tag{1}$$

$$q_{\text{radiation}}(T_1, T_2) = \frac{\sigma_{SB}\left(T_1^4 - T_2^4\right)}{\frac{1-\epsilon_1}{\epsilon_1 A_1} + \frac{1}{A_2}\left(\frac{1}{\epsilon_2} + \frac{2N_{\text{MLI}}}{\epsilon_{\text{MLI}}} - N_{\text{MLI}}\right)}, \tag{2}$$

$$q_{\text{contact}}(T_1, T_2) = \int_{T_2}^{T_1} C_{\text{contact}}(T)\, dT. \tag{3}$$

In the above equations, $A_{CS}$ is cross section area, $L$ is longitudinal separation between the temperature nodes, $k(T)$ is temperature dependent thermal conductivity, $\sigma_{SB}$ is Stefan-Boltzmann constant, $\varepsilon$, $A$, and $N_{MLI}$ are emissivity, surface area, and number of MLI layers, and $C_{contact}$ is contact conductance.

The expressions given by equations (1)-(3) can be applied to each temperature node (cryocooler thermal stages $T_{CC1}$ and $T_{CC2}$, thermal shield $T_{TS}$, magnet $T_M$, and the temperatures upstream and downstream of the pressed contact $T_{L,U}$ and $T_{L,D}$) in figure 2 to write down the energy conservation equation for that node. The result is a first order ordinary differential equation (ODE). Combining the ODE for each note results into a system of ODE given by equation (4):

$$\left(\sum mc_p(T_i)\right)\frac{dT_i}{dt} = \left(\sum q_{\text{in}}\right)_i - \left(\sum q_{\text{out}}\right)_i \qquad i = [\text{TS}, \text{CC2}, \text{M}, \text{LU}, \text{LD}, \text{CC1}] \tag{4}$$

In equation (4), the left hand side represents the rate of thermal energy removal from a node while the right hand side represents the net thermal energy flow through this node at a given time, $t$. The quantity q is a combination of the heat transfer channels given by equations (1)-(3).

The solution of the ODE system given by equation (4) yields the temperature vs. time curve for each node. The solution requires numerical integration because of the strong temperature dependence of material properties used to construct the system. The ODE system is solved in Python using the in-built *odeint* library. Figure 3 shows sample cooldown profiles of the six temperature nodes defined in figure 2. More details such as the cryocooler capacity, contact resistance, the properties of materials, etc. can be discerned by the accompanying Python script.

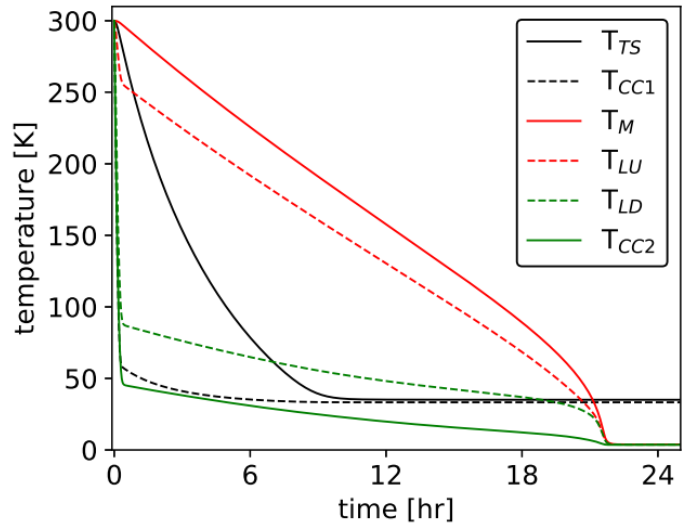

Figure 3. Sample cooldown profiles for the temperature nodes in figure 2.

## 4. Summary

This note presents a numerical framework for calculating the cooldown characteristics of a cryocooler conduction-cooled device. The framework is deployed to calculated cooldown profiles and time of a small superconducting magnet conduction-cooled to two-stage cryocoolers. A system of ODEs is first developed by applying the principle of conservation of energy to various internal parts of the magnet system. The ODE system is solved in Python to yield the cooldown profiles. The complete Python script is provided for the benefit of the reader who can plug in the design parameters of their own systems and calculate the cooldown characteristics. Furthermore, a parametric sweep can be implemented using the script for cooldown time optimization of he system.

**Acknowledgement**

This work was supported by Fermi Research Alliance, LLC under Contract No. DE-AC02-07CH11359 with the U.S. Department of Energy, Office of Science, Office of High Energy Physics.

**FERMILAB-TM-2782-TD**